\documentclass{mem}
\usepackage{natbib}\usepackage{txfonts}\usepackage{balance}
\usepackage{graphicx}
\usepackage[a4paper,breaklinks,dvipdfm]{hyperref}
\idline{75}{282}
\begin{document}
\def\teff{$T\rm_{eff }$}
\def\kms{$\mathrm {km s}^{-1}$}
\def\HI{\hbox{H~$\scriptstyle\rm I\ $}} 
\def\TH{\hbox{\textsc{THESEUS}~}}

\title{
First Stars, Reionization and Gamma-Ray Bursts
}

   \subtitle{}

\author{A. Ferrara}

\institute{
Scuola Normale Superiore, Piazza dei Cavalieri 7, I-56126 Pisa, Italy
}

\authorrunning{Ferrara}

\titlerunning{First Stars, Reionization and Gamma-Ray Bursts}

\abstract{Gamma-Ray Bursts represent unique tools to study the early phases of cosmic evolution, the formation of the first stars and galaxies. Absorption line spectra of these sources located in the Epoch of Reionization might provide us with key information about these remote, and yet fundamental, stages of cosmic history. I will briefly review these issues, highlighting the immense leap that the \TH mission could represent for finding answers to the many fundamental open questions. 

\keywords{cosmology: large scale structure - intergalactic medium - quasars: absorption
lines - gamma-ray: bursts - galaxies: high redshift - luminosity function}
}
\maketitle{}

\section{Introduction}
Right after the recombination epoch, approximately located at redshift $z \simeq 1000$, the Universe entered a phase called the ``Dark Ages'', where no significant radiation sources existed. The cosmic gas, essentially a mixture of hydrogen and helium produced during the first three minutes, was largely neutral at this stage. The small inhomogeneities in the dark matter density field present during the recombination epoch started growing via gravitational instability giving rise to highly nonlinear structures, i.e. collapsed haloes, within which the first stars could form.  The collapsed haloes form potential wells whose depth depends on their mass and the baryons then “fall” in these wells. If the mass of the halo is high enough (i.e., the potential well is deep enough), the gas dissipates its energy, cools via atomic or molecular transitions and fragments within the halo. This produces conditions appropriate for the formation of the first stars and galaxies. Once these luminous objects form, the Dark Ages are over. 
The first population of luminous stars and galaxies generate ultraviolet (UV) radiation through nuclear reactions. In addition to galaxies. The UV radiation contains photons with energies $> 13.6$ eV which are then able to ionize hydrogen atoms in the surrounding medium, a process known as reionization. Reionization is thus the second major change in the ionization state of hydrogen (and helium) in the Universe (the first being recombination). 

The ideal technique to study the evolution of cosmic reionization is the measurement of the 21 cm from neutral hydrogen atoms. 
The 21 cm line is associated with the hyperfine transition (often referred to as the "spin-flip" transition, as the proton and electron spins go from parallel to anti-parallel) between the triplet and the singlet levels of the neutral hydrogen ground state.
The 21 cm sky contains fluctuations around the mean (``global'') signal. This fact is in almost perfect analogy with the more familiar case of the CMB radiation, which on top of the mean intensity given by the 3 K black body radiation, shows small angular fluctuations on a wide range of scales. These fluctuations encode information on the physical state of matter at the recombination epoch. Similarly, the 21 cm fluctuations, once observed, will tell us about the physical state of hydrogen, largely representative of all baryons, in the Dark Ages and in the Epoch of Reionization (EoR).
%
\begin{figure}
\resizebox{\hsize}{!}{\includegraphics[clip=true]{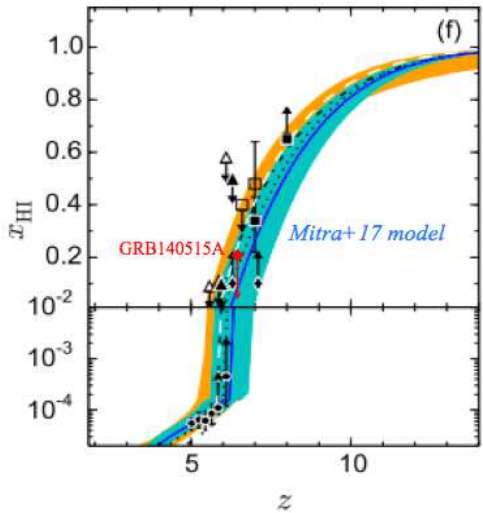}}
\caption{\small Redshift evolution of the cosmic neutral hydrogen fraction from the data constrained model by Mitra, Choudhury \& Ferrara (2018) including available constrains from quasar and GRB spectra. The data point relative to GRB 140515A at $z=6.33$ is shown by the red point.  
}
\label{Fig01}
\end{figure}

At high redshifts, when all the IGM is essentially neutral, the spectrum closely resembles the underlying dark matter spectrum. However, as ionized bubbles appear they boost the power on intermediate/large scales corresponding to their own sizes (several Mpc). At the same time small scales are depressed with respect to the dark matter spectrum: this is because within ionized bubbles the ionized fraction is uncorrelated with the small-scale density perturbations.  These processes imprint a characteristic S-shape to the (nondimensional) power spectrum with the normalization decreasing as redshift and neutral fraction decrease.
Unfortunately, detecting the signal represents a formidable challenge, as foregrounds in the low-frequency radio band around 100 MHz are much stronger than the signal and removing them is far from trivial. The promise is that radio telescopes as SKA1 and, perhaps, HERA will be able to eventually achieve this goal. 

In the meantime reionization continues to be studied using the canonical tools: the polarization of the CMB, and most relevant here, absorption line spectra of bright sources, typically quasars,  located in the reionization epoch. However, the paucity of these objects at early epochs has prevented systematic studies of the EoR. Long gamma-ray bursts (GRB) may constitute a complementary way to
study the reionization process avoiding the proximity effects and probing higher redshifts. This has been made possible for the first time after the detection of GRBs at $z >5$ (Gehrels et al. 2004), and specifically demonstrated by the studies by Totani et al. (2006), and Gallerani et al. (2008) who have used this object to constrain the ionization state of the intergalactic medium (IGM) at high redshift by modelling its optical afterglow spectrum. 

In addition to comic reionization, GRBs can be used also to probe the star formation in early galaxies (Lamb \& Reichart 2000; Prochaska et al. 2007) potentially to higher redshifts (e.g. GRB 090429B at $z = 9.4$, Cucchiara et al. 2011)  than allowed by galaxies alone. Useful information can be deduced on the physical properties of the GRB host galaxy and also on the nature of the progenitor, thus ideally allowing to identify potential sites of the first metal-free (Pop III) episodes of star formation. In what follows we concentrate on these aspects, elucidating the potentiality of GRB studies for reionization and early galaxy 
formation studies. 

\section{Studying reionization with GRBs}
One of the most striking applications of the absorption line technique to high redshift GRB is the one provided by the study by Chornok et al. (2014).  These authors took spectra of the optical afterglow of the GRB 140515A located at $z\approx 6.33$. Apart from a small transmissivity window the flux shortwards of the Ly$\alpha$ line is totally blocked by the Gunn-Peterson through, indicative of a relatively low-ionization line of sight. The red damping wing of the Ly$\alpha$ lines can be modelled making three different assumptions: (a) a single host galaxy absorber at $z=6.327$ with $\log(N_{HI})=18.62 \pm 0.08$; (b) pure IGM absorption from $z=6.0$ to $z=6.328$ with a constant neutral hydrogen fraction of $x_{HI}=0.056^{+0.011}_{-0.027}$; and (c) a hybrid model with a host absorber located within an ionized bubble of radius 10 comoving Mpc in an IGM with $x_{HI}=0.12 \pm 0.05$. Regardless of the model, the sharpness of the dropoff in transmission is inconsistent with a substantial neutral fraction in the IGM at this redshift. Thus, an accurate modelling of the Ly$\alpha$ profile can provide crucial information on the state of the IGM in the EoR. Performing this experiment on a large number of GRBs at high-$z$ is crucial in order to gather a statistically significant measurement of the mean neutral fraction as a function of cosmic time. In addition, spatial variations of $x_{HI}$ might allow to unveil the clustering of the ionizations field and of the sources. 
%
\begin{figure}
\resizebox{\hsize}{!}{\includegraphics[clip=true]{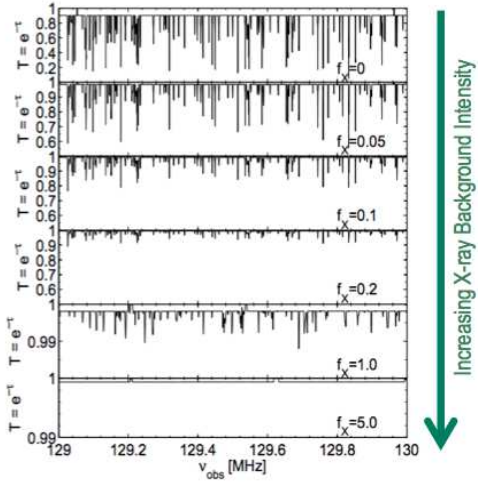}}
\caption{\small  Relative transmission along a line of sight to a GRB afterglow at redshift $z =10$ from which the 21 cm forest
spectrum is clearly seen. The six panels from top to bottom show the results for increasing intensities of the X-ray background,
parameterized by the variable $f_X$ (see Xu et al. 2011) for details. Note that the y-axes
are different between panels.
}
\label{Fig02}
\end{figure}
Results as the above one might considerably help constraining models of cosmic reionization. Mitra et al. (2018), extended their  previous MCMC-based data-constrained semi-analytic reionization model to include also study the role of quasars on global
reionization history. The data seems to favour a standard two-component picture where quasar contribution become negligible at $z > 6$ and a non-zero escape fraction of  about 10\% is needed from early-epoch galaxies. For such models, the mean neutral hydrogen fraction decreases to $x_{HI} \approx 10^{-4}$ at $z = 6.2$ from about 0.8 at $z = 10$. In addition helium becomes doubly
ionized at much later time, $z=3$. These findings are fully consistent (see \ref{Fig01}) with the results from GRB 140515A, thus supporting the model. \TH, by detecting a large number of $z>6$ GRB, where only a handful of quasars are known, will allow to push these type of studies well into the cosmic dawn at $z>10$.

\begin{figure}
\resizebox{\hsize}{!}{\includegraphics[clip=true]{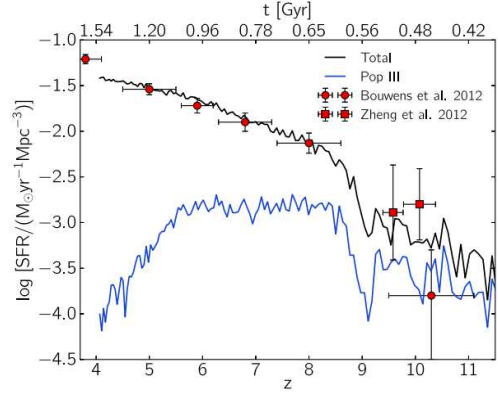}}
\caption{\small  Cosmic star formation rate density (SFR) as a function of redshift (age of the Universe) for all stellar populations (black line) and for PopIII stars only (blue line). Data points (in red) are taken from Bouwens et al. (2012) and Zheng et al. (2012).
The figure is taken from Pallottini et al. (2014x).
}
\label{Fig03}
\end{figure}

In a similar spirit, Gallerani et al. 2008 pioneered the so-called ``Dark Gaps'' technique as a statistical tool to use GRB spectra to investigate reionization. In practice, the spectra of high-redshift sources (as QSOs and GRBs) bluewards of the Ly$\alpha$ are characterized by dark portions (gaps) produced by intervening neutral hydrogen along the line of sight. The statistics of the dark gaps in GRB absorption spectra represent exquisite tools to discriminate among different reionization models. These authors compute the probability to find the largest gap in a given width range $[W_{max}, W_{max} + dW]$ at a flux threshold $F_{th}$ for burst afterglows at redshifts $6.3 < z < 6.7$. Different reionization scenarios populate the $(W_{max}, F_{th})$ plane in a very different
way, allowing to distinguish among different reionization histories. When applied to a specific high-$z$ ($z=6.29$) GRB 050904, Gallerani et al. where able to show that the observation of this burst strongly favors reionization models predicting a highly
ionized IGM ($x_{HI} = 6.4 \pm 0.3 \times 10^{-5}$), again consistent with the models presented in Fig. \ref{Fig01}.

A perhaps even more exciting possibility is provided by 21 cm radio observations of GRB afterglows. This idea has been first explored by Xu et al. (2011) who investigated the 21 cm absorption lines (known as the ``21 cm forest'') produced by non-linear structures during the early stage of reionization, i.e. the starless minihaloes and the dwarf galaxies. The infalling gas velocity around minihaloes/dwarf galaxies strongly affects the line shape and, with the low spin temperatures outside the virial radii of the systems, gives rise to horn-like line profiles. The authors compute synthetic spectra of 21 cm forest for the radio afterglows of gamma-ray bursts (GRBs). Broadband observation against GRB afterglows can also be used to reveal the evolving 21 cm signal from both minihaloes and dwarf galaxies. The number, strength and clustering of 21 cm absorption lines depend very sensitively on the intensity of the X-ray background produced by e.g. high-mass X-ray binaries or early black holes (as shown in Fig. \ref{Fig02}). The experiment would then offer a unique opportunity to explore cosmic dawn and the rise of the first structures, including black holes.  Again, the synergy between \TH and SKA will be fundamental in order to achieve these ambitious goals.

\section{First stars and metal enrichment}
Elements heavier than Helium were not formed during Big Bang Nucleosythesis and they had to await the formation of the first stars to be produced in their interiors. As a result, the first cosmic star formation activity had to occur in a primordial composition gas. Stars formed in these physical conditions are usually dubbed as Pop III stars. Theoretical studies predict that they should be typically massive (albeit precise predictions on their masses are currently lacking: for recent reviews on the subject see, e.g. Ferrara 2012, Bromm 2013). This is due to the combined effect of the larger gas fragmentation scale and accretion rate, and the very limited opacity. On the other hand, observations of present-day stellar populations (Pop II/I stars) show that stars form according to a Salpeter Initial Mass Function (IMF) with a characteristic mass of $1 M_\odot$, below which the IMF flattens. Thus, unless the current picture of primordial star formation is lacking in some fundamental ingredient, a transition between these two modes of star formation must have occurred at some time during cosmic evolution. The emerging physical interpretation, initially worked out by Schneider et al. (2002), states that the fragmentation properties of the collapsing clouds change as the mean metallicity of the gas increases above a critical threshold, $Z_{cr} = 10^{-5\pm 1} Z_\odot$, above which enhanced cooling provided by metals and dust should induce rapid fragmentation of the birth cloud in solar-mass sub-units.

\begin{figure*}
\resizebox{\hsize}{!}{\includegraphics[clip=true]{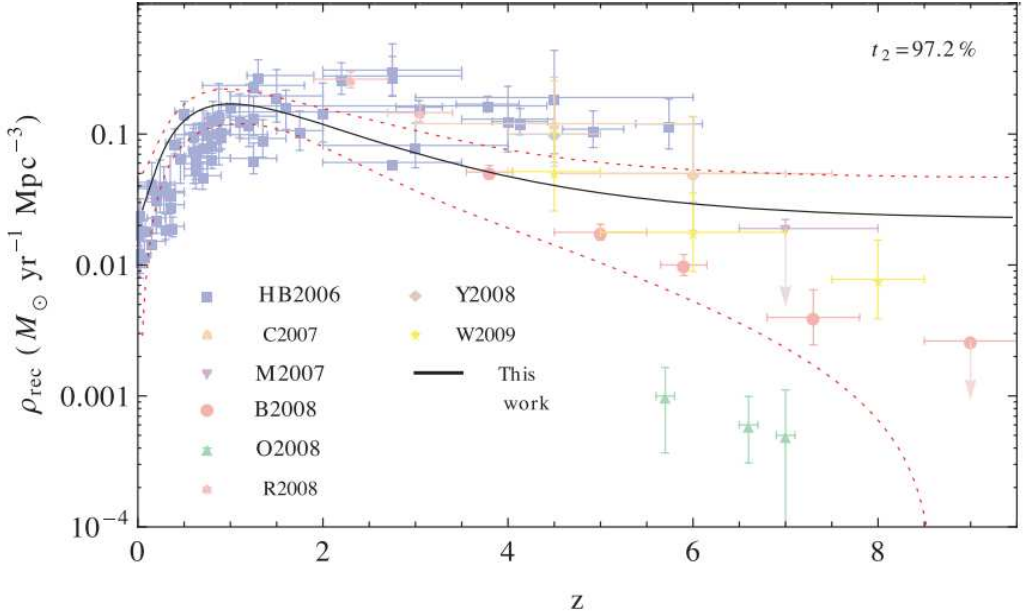}}
\caption{\small  PCA reconstruction from Swift data using two PCs, compared with
independent SFH determinations (light points, not used in our calculations).
The black solid line is the PCA best-fitting reconstruction using two PCs;
the red dashed lines correspond to 2$\sigma$ confidence levels. The inset shows
the cumulative percentage of total variance. Taken from Ishida et al. (2011).
}
\label{Fig04}
\end{figure*}

Tornatore, Ferrara \& Schneider (2007) applied this physical argument on cosmological scales to understand the global Pop III $\rightarrow$ Pop II/I transition during cosmic evolution. They performed numerical simulations following the evolution, metal enrichment and energy deposition of both Pop III and Pop II stars. The most striking result is that they were able to show that, due to inefficient heavy element transport by outflows and slow transmission during hierarchical growth, Pop III star formation can continue at least down to low redshifts ($z \approx 2$), albeit at a rate which is about 1/1000 of the Pop II star formation at that epoch.  They also noted that Pop III star formation proceeds in a ”inside-out” mode in which formation sites are progressively confined at the periphery of collapsed structures, where the low gas density and correspondingly long free-fall timescales result in a very inefficient astration. The above findings have been confirmed by more modern simulations (Fig. \ref{Fig03}) studies (Pallottini et al. 2014, Muratov et al. 2015, Jaacks et al. 2018). These results have important implications for the search of the so far unsuccessful searches for the elusive Pop III stars; in addition they open very interesting perspectives for the use of high-redshift GRBs to pin-point events arising from the final stages of Pop III stars evolution. Among these are that at earlier times on should expect: (a) an increasing fraction of Pop III-dominated galaxies; (ii) a larger frequency of core-collapse and pair-instability supernova events; and, most importantly here, (c) an increasing rate of Pop III GRBs.

\begin{figure*}
\resizebox{\hsize}{!}{\includegraphics[clip=true]{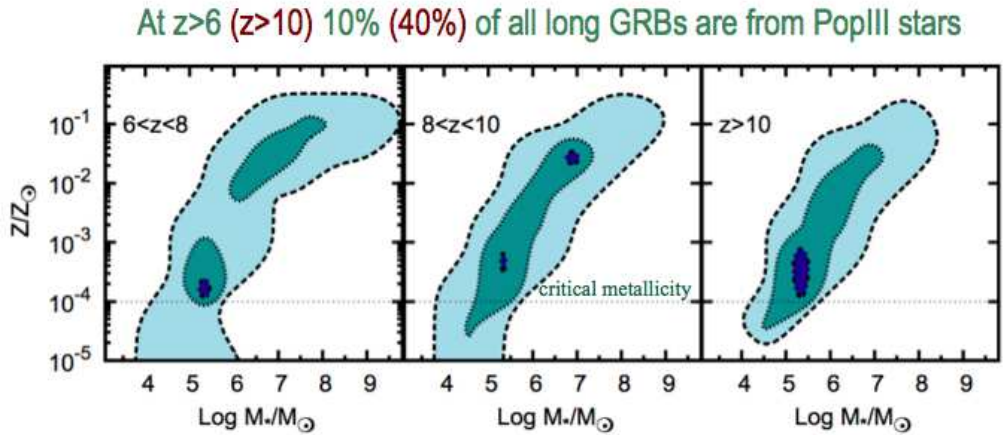}}
\caption{\small  Probability of having a Pop III GRB  in a host galaxy with given stellar mass and metallicity, in three different redshift bins. The contours refer to a probability of 100\% (outermost contour), 75\% and 25\% (innermost). The horizontal line indicates
the critical metallicity, $Z_{cr}$. Adapted from Campisi et al. (2011).
}
\label{Fig05}
\end{figure*}

Due to their high luminosity, GRBs can be used as SFH probes into the very distant Universe as it was first suggested by Lamb \& Reichart (2000). In principle, the redshift distribution of GRBs can give us important clues on the early stages of cosmic history. In practice, the connection between GRBs and the underlying host galaxy star formation model is often complicated, thus making the probe value subject to uncertainties. To circumvent this problem, Ishida et al. (2011) pioneered a novel method based on the Principal Component Analysis (PCA). The main advantage of such approach is to avoid the necessity of assuming an ad hoc parametrization
of the star formation history (SFH). The method can be robustly validated by reconstructing a known SFH from Monte Carlo-generated mock data, applied to the available Swift data of GRBs with known redshift, and finally compared against the SFH obtained by independent methods. The results are striking. The combination of GRB data and PCA suggest that the level of star
formation activity at $z = 9.4$ could have been already as high as the present-day one ($\approx 0.01 M_\odot \mathrm{{yr}}^{-1} \mathrm{{ Mpc}}^{-3}$). This is a factor 3-5 times higher (see Fig. \ref{Fig04}) than deduced from high-$z$ galaxy searches (Oesch et al. 2017) through drop-out techniques. Such discrepancy might indicate that a sizable fraction of star formation during EoR occurs within dust-obscured regions  and/or is hosted in galaxies to faint to be detected by current surveys. Interestingly, reionization might proceed in a very different manner depending on the balance between the two above hypothesis as in the dust-obscured case ionizing photons would be degraded into far infrared thermal emission before escaping from the parent galaxy. Alternatively,  if star formation connected with the observed GRBs is located in very faint galaxies,  this might alleviate the longstanding problem of a photon-starving
reionization.

A quantitative analysis of the detectability of Pop III GRBs has been carried out by Campisi et al. (2011). By using cosmological simulations similar to those by Tornatore et al. (2007), they analyzed the correlation between early Pop III stars and long GRBs. 
From the Swift observed rate of long GRBs, and assuming that as of today no GRB event has been associated to a PopIII star, they conclude that the upper limit for the fraction of long GRBs produced by PopIII stars is in the range $0.006< f_{GRB} <0.022$. Once a detection threshold compatible with the BAT instrument is applied, Campisi and collaborators conclude that the expected fraction of PopIII GRBs is about 10\% of the full  long GRB population at $z >6$, becoming as high as 40\% at $z >10$ (Fig. \ref{Fig05}). If such prediction will be confirmed by \TH a new, unique tool will become available to investigate how star formation started in the Universe.

Detecting Pop III GRBs might also allow to study their host galaxy, and hence discriminate between the two hypothesis discussed above on the alleged dearth of high redshift star formation missing from the current surveys (but badly needed to explain reionization).  From their simulations Campisi et al. found that the average metallicity of the galaxies hosting a Pop III GRB is typically higher than the critical metallicity, $Z_{cr}$, for the transition between Pop III and Pop II stars. This is due to the rapid enrichment of the galaxy interstellar medium triggered by highly energetic explosions of the first stars. Interestingly, they also concluded that, independently on redshift, the probability of finding a Pop III GRB peaks for low-mass (stellar mass $< 10^7 M_\odot$) galaxies. 
This is a consequence of the fact that these systems are more likely contain unpolluted patches of gas. 

Detecting Pop III GRBs might shed light also on the IMF of the first stars based on the metal abundance ratios measured in high-z GRB afterglow observations (Ma et al. 2017). Although the results might be dependent on the accuracy of the stellar yields, for some class of models involving the explosion of massive or very massive SN models, the abundance ratio distribution of the host galaxies  retains memory of the stellar IMF, allowing to broadly distinguish among the different theoretical predictions for this quantity. 

Finally, GRB absorption studies may also be used to infer the spatial distribution of metal enrichment of early galaxies and their circumgalactic medium. In their application to GRB 140515A, Chornok et al. (2014) found no narrow absorption lines from the host galaxy, indicating a host metallicity of $z < 0.16\, Z_\odot$; in addition, if all of the hydrogen absorption is due to the host galaxy, the unusually low column density for a GRB sightline might show a possible evidence of a high escape fraction of ionizing photons from this system. 

\section{Final remarks}
In addition to representing a fascinating physical phenomenon, GRBs are one the most promising tools to study the early stages of cosmic evolution. So far we have just scratched the surface of such epoch which represents the current cosmic frontier. The advent of new breakthrough facilities like JWST, ELTs, SKA and most importantly in this context, THESEUS, may truly revolutionize our understanding of first stars and galaxy formation, along with the many unknowns still associated with the reionization process. Such mission will provide answers to long-standing key questions. Among these are: (i) Do Pop III GRB exist and can we use them to probe the first generation of stars? This question impinges on the shape of the primordial IMF of these stars, and on their angular momentum properties; (ii) What is the fraction of Pop III GRBs and how does it evolve with redshift? (iii) How do we exploit at best GRB absorption line experiments to study cosmic reionization and metal enrichment of the IGM, and how can we maximize the synergy with other space-born and ground facilities? The answer to these questions might bring genuine surprises as we continue the exploration of the most remote regions of our Universe.

\begin{acknowledgements}
AF acknowledges support from the ERC Advanced Grant INTERSTELLAR H2020/740120
\end{acknowledgements}

\bibliographystyle{aa}

\end{document}